Properties of Electrodeposited Molybdenum Disulfide on Zinc Oxide and Zinc Oxide/Zinc Sulfide Nanowires

Lee Kendall, Dawn Ford, Giovanni Zangari, Stephen McDonnell

Department of Materials Science and Engineering, University of Virginia, Charlottesville, VA 22904

*Corresponding author: mcdonnell@virginia.edu

Abstract:

We developed a facile and scalable 3-step hydrothermal, electrodeposition, and annealing technique to synthesize a variety of nanowire heterostructures. The heterojunction catalysts of CP-ZnO-MoS$_2$ and CP-ZnO-ZnS-MoS$_2$ both saw an increase in catalytic activity in the acidic regime, with overpotentials to reach 10 mA/cm$^2$ of 181 mV and 154 mV respectively, over their constituent materials. The CP-ZnO-ZnS-MoS$_2$ catalysts also saw an increase in catalytic activity in the alkaline conditions, requiring only 209 mV to reach 10 mA/cm$^2$. This has been attributed primarily to the synergistic properties of the band alignment structure and the increased surface area afforded by the nanowire structure. The 1T phase was also noted in the CP-ZnO-MoS$_2$ sample, suggesting that this also played a role in increasing the catalytic activity of the sample. Overall, the synthesized nanowire heterostructures were found to be high-performing and gives insight into how the changing band alignment and morphology can play a significant role in increasing the catalytic performance towards HER.

Introduction:

Recently, hybrid core/sheath nanostructures consisting of a metal oxide semiconductor core (ZnO, TiO$_2$, MnO, etc.) covered by transition metal dichalcogenides (TMDC) layers and other chalcogenides, such as ZnSe, CdS, and CdSe, have been explored as potential efficient catalysts due to their more efficient charge separation.[1–5] Among the metal oxides, ZnO is an important group II-IV semiconductor with a band gap of 3.37 eV and a Hall mobility on the order of 200 cm$^2$V$^{-1}$s$^{-1}$. Its low cost, non-toxicity, ease of synthesis and crystallization, and natural abundance make it an ideal and promising material for catalytic applications, most notably as a photocatalyst. It is currently regarded as the best alternative to TiO$_2$ for electrochemical applications due to its faster charge transfer and prolonged electron lifetime when compared to TiO$_2$. However, ZnO by itself is inactive for HER and suffers from degradation in both acidic and alkaline conditions due to its amphoteric nature. As such, it is often coated with an active species that can be enhanced by the ZnO while also serving to protect it.

The ZnO/MoS$_2$ and other ZnO/TMDC interfaces have been studied previously and have shown promise given a reduction in band gap, maintained charge mobility, and improved charge separation due to the lattice mismatch. These studies have been accomplished with a variety of methods, from drop-casting, solvothermal, and metal-organic chemical vapor deposition[6–11] with a variety of morphologies, from films, nanoparticles, nanorods, and nanowires.[8,11–13] However, these studies have all been conducted on flat, 2D substrates, limiting their overall active surface area. As such, our study will be conducted on a 3D, porous substrate to maximize the overall active surface area.

Taking this a step further, it is possible to elevate the catalytic activity via charge movement engineering by incorporating an additional semiconductor with the appropriate band structure and forming a ternary heterostructure. When done, the electrons can be further spatially confined through the creation of electron and hole sinks, further enhancing charge separation.[14,15] Furthermore, the ternary heterostructure generates an additional interface with further lattice mismatch, modifying the band structure and leading to increased catalytic activity.[16]

Finally, the impact of band alignment will be explored. The two primary types of band alignments that will be explored are type-I (straddling) and type-II (staggered) and can be seen in **Error! Reference source not found.**. In a type-I band alignment, the conduction band minimum of SC-A is higher than that of SC-B while the valence band maximum of SC-A is lower than that of SC-B. Therefore, both electrons and holes from SC-A are transported to the conduction band and valence band of SC-B, which does not facilitate charge separation. Type-II differs in that the electrons from SC-A are transferred to SC-B whereas the holes from SC-B are transferred to SC-A, resulting in better charge separation.[17]

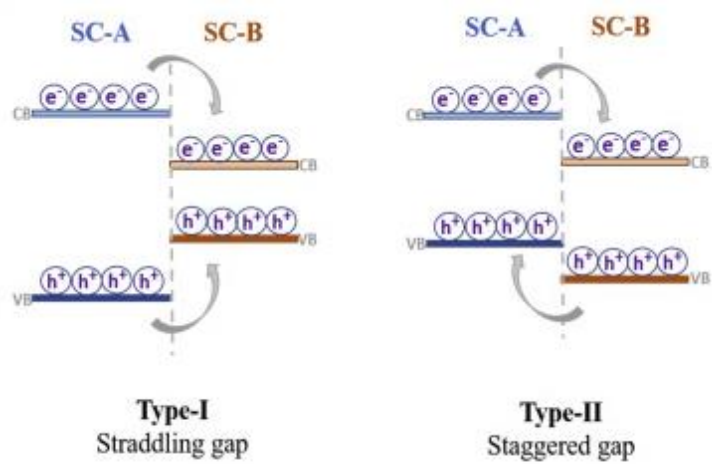

Figure 1: Different types of band alignment in a heterojunction reprinted with permission from Arotiba et al. Current Opinion in Electrochemistry 22 (2020) 25 Elsevier [17]

In the $ZnO/MoS_2$ and the $ZnO/ZnS/MoS_2$ heterostructure, both type-I and type-II band alignments have been reported. However, as mentioned earlier, there is a lack of consensus on the location of the relative band positions depending on synthesis and characterization technique as can be seen in Figure 2.[6,18–22] There is a variety of type-I and type-II alignments for the same heterojunction, both between ZnO/ZnS, $ZnO/MoS_2$, and $ZnS/MoS_2$. To this end, differences in band structure, even with systems with identical constituent materials synthesized in similar manners, are quite common. This is particularly true for materials with conduction and/or valence bands located at a similar level. Additionally, this is further compounded by the fact that the measured band structure is extremely sensitive to, among others, the density of defects, presence of doping, and technique used to measure the band alignment (XPS, impedance-based measurements, photoluminescence, etc.). A change in synthesis conditions, synthesis methods, and choice of substrate is likely to impact the band structure and alignment of the materials.

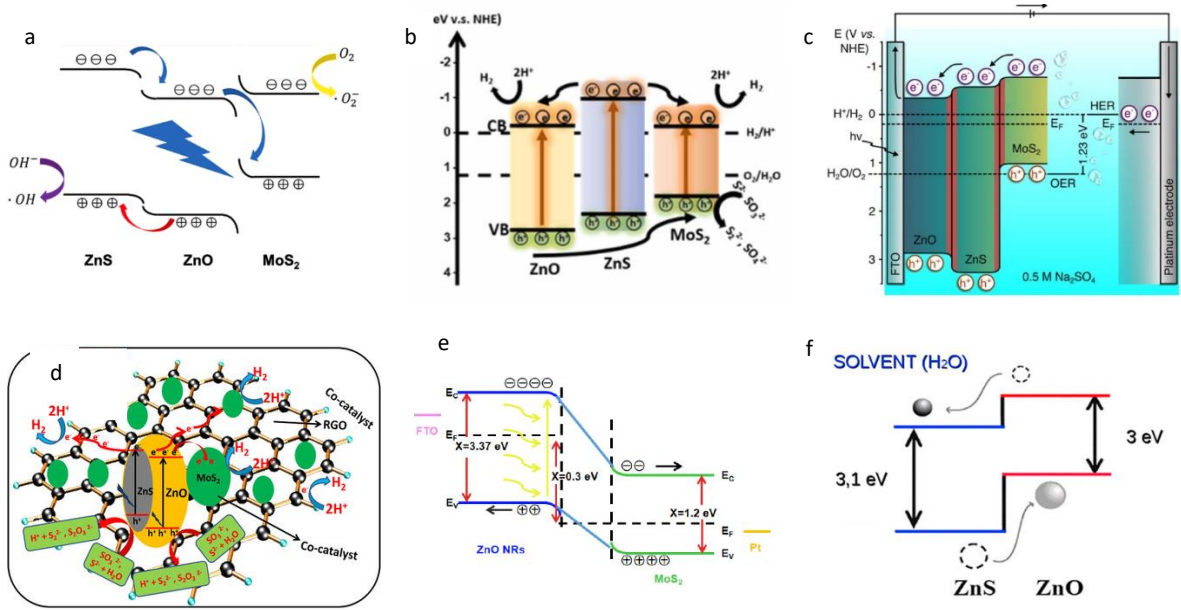

Figure 2: Energy band diagrams of a) ZnO/ZnS/MoS$_2$ piezoelectric nanoarrays reprinted with permission from Lee et al. Nano Energy 93 (2022) 106867 Elsevier [6] b) ZnO/ZnS/MoS$_2$ heterostructures reprinted with permission from Fu et al. Applied Surface Science 556 (2021) 149695 Elsevier [22] c) ZnO/ZnS/MoS$_2$ nanostructure reprinted with permission from Fareza et al. ACS Applied Nano Materials 5 (2022) 16051 American Chemical Society [18] d) ZnS-ZnO-MoS$_2$ heterostructure supported on RGO reprinted with permission from Krishnan et al. ChemSusChem 10 (2017) 3588 John Wiley & Son [19] e) ZnO@MoS$_2$ nanocomposite reproduce from Wu et al Micromachines 2020, 11(2), 189 MDPI [21] f) ZnO@ZnS nanostructures reprinted with permission from Flores et al. Materials Chemistry and Physics 173 (2016) 347 Elsevier [20]

### 3.1.2 Synthesis Methods
### 3.1.2.1 Synthesis of ZnO Nanowires

To synthesize ZnO nanowires (NWs), a simple hydrothermal technique by Wang et al. [update citation] was used. The sequence of process steps to create the heterostructures is schematically illustrated in Figure 3. Carbon paper (CP) was heated to 500 °C in an MTI OTF1200X single-zone furnace for 1 hour in air to increase its wettability. A seed layer was formed by soaking the CP in an aqueous solution containing 0.1M potassium permanganate (KMnO$_4$) for 1 hour to form a seed layer. Then, this seeded CP was placed into an autoclave with a solution containing 15 mM zinc nitrate hexahydrate (Zn(NO$_3$)$_2$)*6H$_2$O), 15 mM hexamethylenetetramine ((CH$_2$)$_6$N$_4$), and 0.4 parts ammonia (e.g. 100 mL solution with 96 mL H$_2$O and 4 mL ammonia). The autoclave was then sealed and placed into the MTI

Compact Forced Air Oven at 90 ℃ for 24 hours. After cooling to room temperature, the CP/ZnO NWs was water washed and dried at 80 ℃ for 3 hours. [23]

### 3.1.2.2 Electrodeposition of MoS$_x$ on ZnO NWs

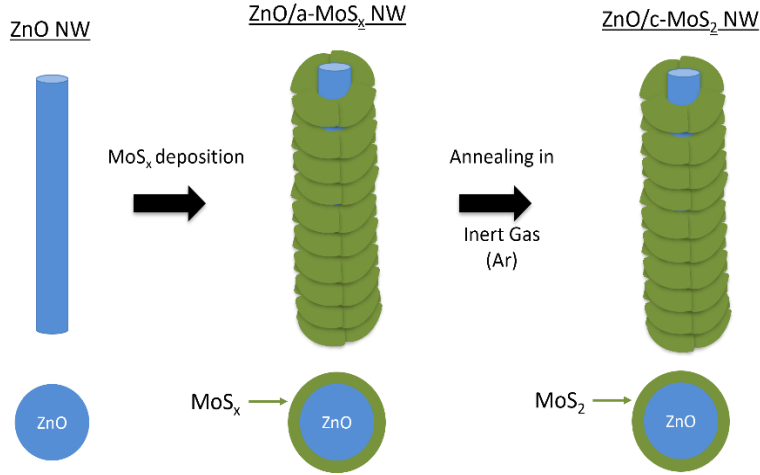

Figure 3: Synthesis steps for CP-ZnO-MoS$_2$

MoS$_x$ was electrodeposited from a chemical bath containing 50 mM sodium molybdate dihydrate (Na$_2$MoO$_4$*2H$_2$O) and 0.2 M sodium sulfide and was based on Levinas et al.'s work.[24] The general equation for the sulfidation of MoO$_4^{2-}$ when titrated from a pH of 13 to a pH of 8 follows

$$MoO_4^{2-} + 4HS^- + 4H^+ <-> MoS_4^{2-} + 4H_2O \qquad (1)$$

The MoS$_4^{2-}$ anion was electrodeposited under potentiostatic conditions at a potential of -1.5 V$_{MSE}$ for 30 minutes with the CP-ZnO as the substrate. It was then removed immediately from the electrolyte and rinsed with DI water. The CP-MoS$_x$ was formed without the presence of ZnO NWs

**Formation of ZnO/MoS$_2$ and ZnO/ZnS/MoS$_2$ nanowires**

To crystallize ZnO-MoS$_x$ into ZnO-MoS$_2$, the MoS$_x$ coated nanowires were annealed in an MTI OTF1200X single zone tube furnace under flowing ultra-pure nitrogen. After initial optimizing, an anneal temperatures of 550 ℃ was selected. To form the ZnO-ZnS-MoS$_2$ nanowires, elemental sulfur was placed upstream to sulfurize the ZnO and to form a ZnS region between ZnO and MoS$_2$. To form ZnO-ZnS, elemental sulfur was placed upstream to sulfurize the ZnO. To form ZnO-MoS$_2$, the ZnO-MoS$_x$ was placed in the furnace, both with and without sulfur upstream.

**Impact of Annealing Temperature**

Figure 4 shows the SEM images of the synthesized CP-ZnO-ZnS-MoS2$_2$ substrates. As it can be seen in Figure 4a, the ZnO NWs are densely packed and are uniformly distributed. These NWs possess a high aspect ratio with diameters in the hundreds of nanometers and lengths of several micrometers. Figure 4b/c demonstrates that following anneals at 450 °C and 550 °C the nanowires structure is retained. Above these temperatures, the nanowire morphology is lost as seen in Figure 4d-f. In addition to the loss of morphology, a small mass loss was also measured via differential scanning calorimetry starting around 550 °C.

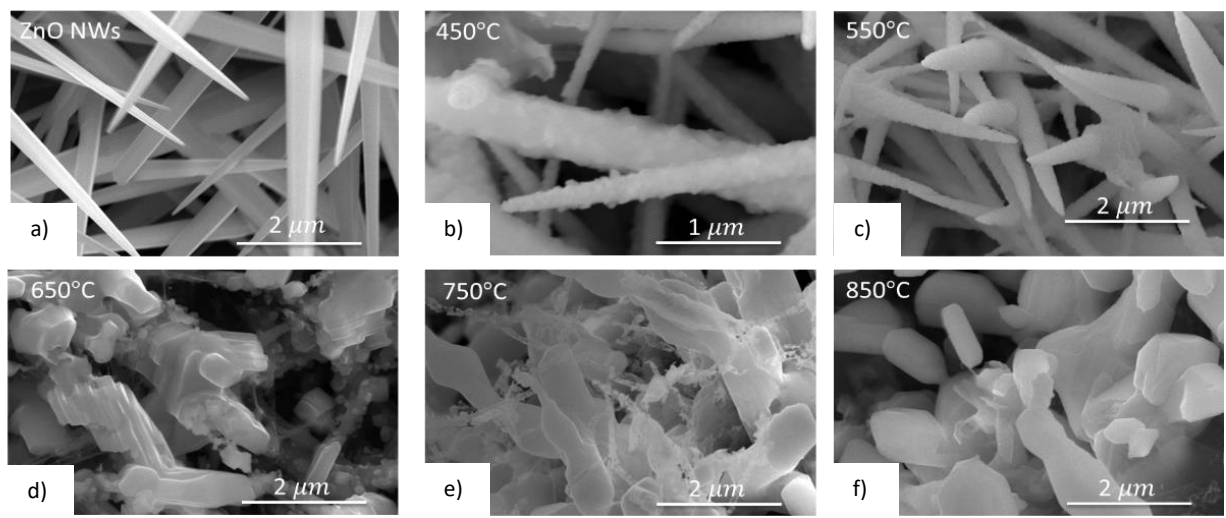

Figure 4: SEM images of a) ZNO NWs, b) ZnO-ZnS-MoS$_2$ annealed at 450 °C, c) ZnO-ZnS-MoS$_2$ annealed at 550 °C, d) ZnO-ZnS-MoS$_2$ annealed at 650 °C, e) ZnO-ZnS-MoS$_2$ annealed at 750 °C, f) ZnO-ZnS-MoS$_2$ annealed at 850 °C

Raman spectroscopy was carried out to understand how the crystal structure evolved as a function of temperature and of sulfur presence while annealing, shown in Figure 5. A brief analysis of the Raman spectrum is given here with a more detailed analysis later in the discussion. The main MoS$_2$ peaks are located at 382 cm$^{-1}$ and 408 cm$^{-1}$ and generally dominate the spectra. The primary ZnS peak is given at 348 cm$^{-1}$ and the primary ZnO is given at 438 cm$^{-1}$ but is generally buried under the broad longitudinal acoustic mode of MoS$_2$ that is located at 450 cm$^{-1}$. Additionally, fluorescence can be seen on the right-hand side of the figure for the samples annealed with sulfur which is due to the presence of the ZnS phase. As such, a few takeaways can be derived from this data. One, the presence of sulfur during annealing encourages the formation of a ZnS phase, which is most pronounced at 550 °C. Without the presence of sulfur no ZnS is formed. This is further confirmed by the lack of fluorescence which is caused by the presence of the ZnS phase. Additionally, this demonstrates that the MoS$_2$ phase is stable at all tested temperatures, even if the ZnO scaffolding is not. Therefore, all samples from here on will be annealed at 550 °C for comparisons.

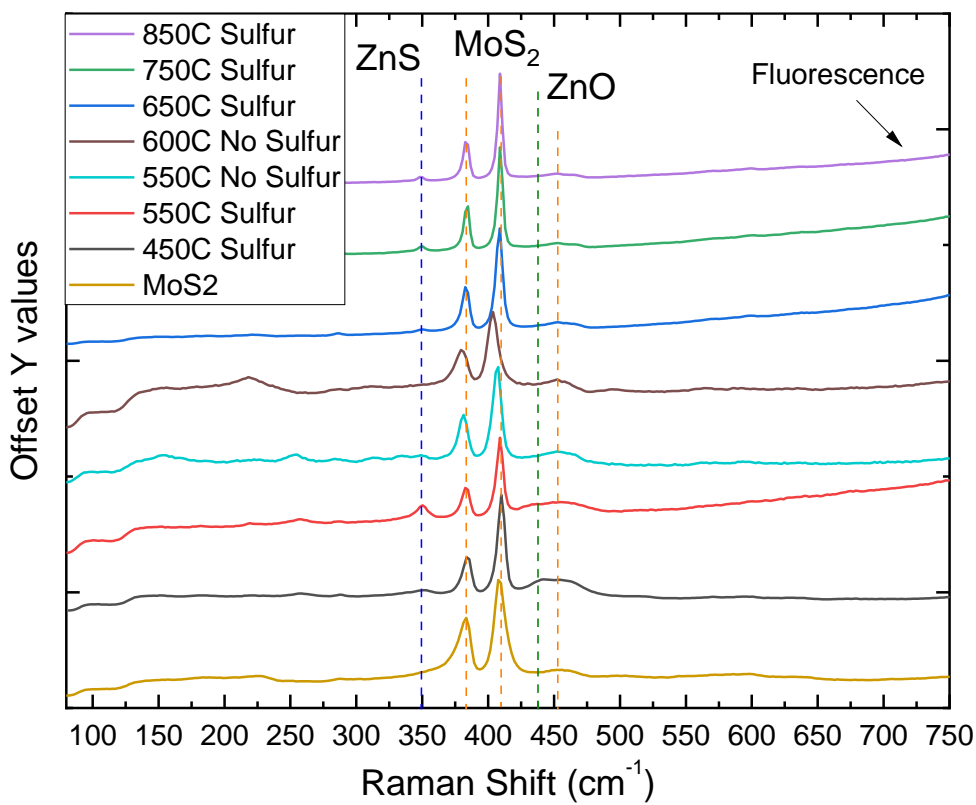

Figure 5: Raman spectra of the CP-ZnO-MoS$_2$/CP-ZnO-ZnS-MoS$_2$ as a function of annealing temperature and presence of sulfur while annealing. The blue dashed lines represent the ZnS phase, the orange lines are attributed to the MoS$_2$ phase, and the green dashed line is attributed to the ZnO phase

**Structural Properties of ZnO/MoS$_2$ and ZnO/ZnS/MoS$_2$**

To compare the impact that annealing in the presence of sulfur has on the morphology of the nanowires, sulfur was placed upstream in the furnace. Figure 6 shows typical SEM images of the various treatments and the well-preserved nanowire structure. The ZnO/ZnS/MoS$_2$ NWs are distinctly more textured than when no sulfurization takes place. This potentially occurs while both ZnO and ZnS are cubic materials, ZnS has a significantly larger lattice parameter than that of ZnO and of hexagonal MoS$_2$.

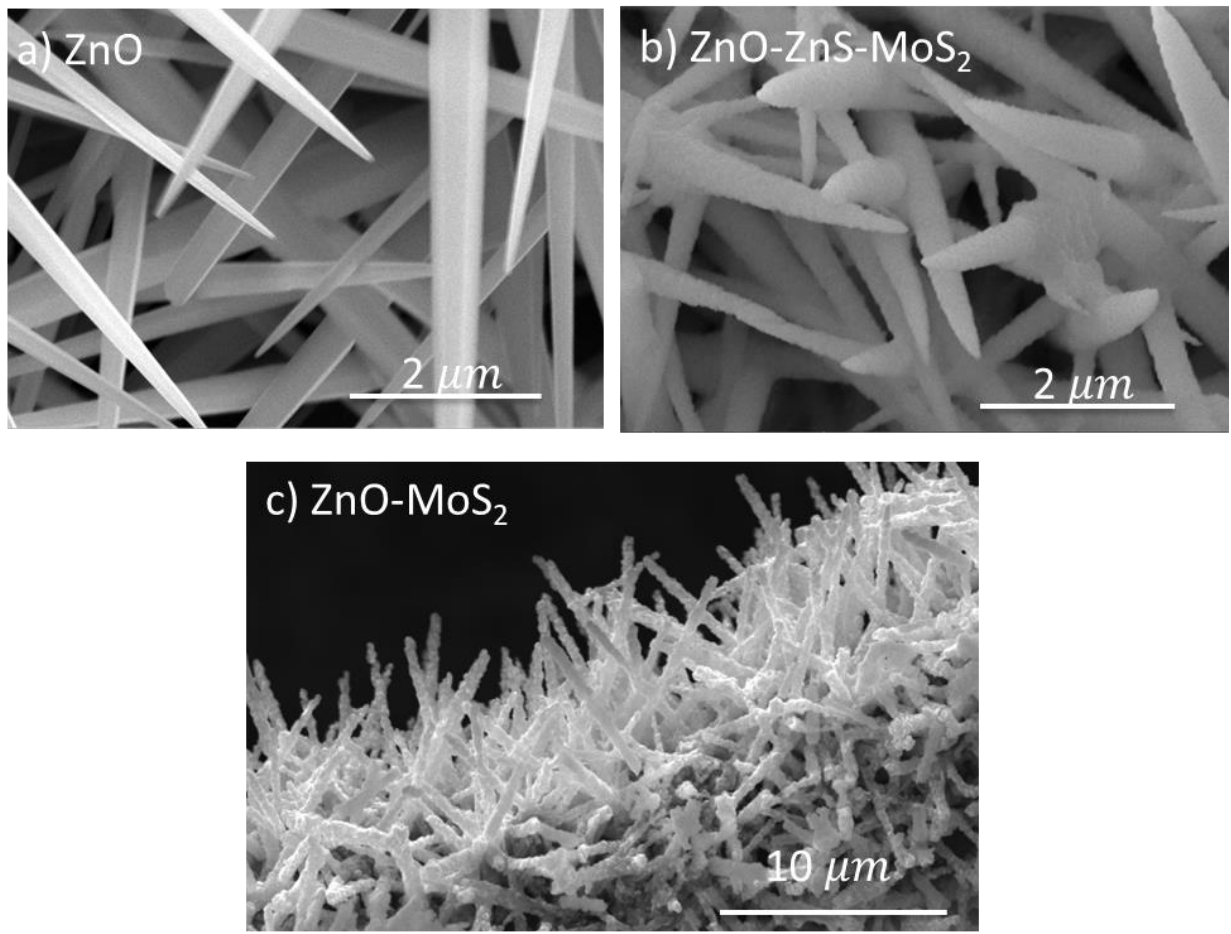

Figure 6: SEM micrographs of a) CP-ZnO NWs, b) CP-ZnO-ZnS-MoS$_2$ NWs, and c) CP-ZnO-MoS$_2$ nanowires

XRD patterns of the various nanowires are shown in Figure 7. The Bragg diffraction hump at 2θ ≈ 14.17° was indexed to the (002) plane of hexagonal MoS$_2$ (ICDD 01-075-1539).[25] The peaks located at 2θ ≈ 17°, 27.8°, 32.7°, 48.66°, 52.25°, and 57.28° are attributed to leftover Na$_2$MoO$_4$ precursor (ICDD 98-004-4523). The peaks located at 2θ ≈ 31.9°, 34.5°, 36.4°, 47.6°, 56.7°, 63.0° are attributed to the cubic ZnO phase (ICDD 01-086-8923). The peaks at 2θ ≈ 27.8°, 28.6°, 30.6°, 43.3°, 47.6°, 51.8°, and 56.5° are attributed to the cubic phase of ZnS (ICDD 04-007-9852). Finally, the peaks at 2θ ≈ 26.6°, 29.5°, and 54.75° are attributed to the substrate, graphic carbon (ICDD 01-075-2078). However, there is a peak at 2θ ≈ 12° and 14.8° that is unidentified.

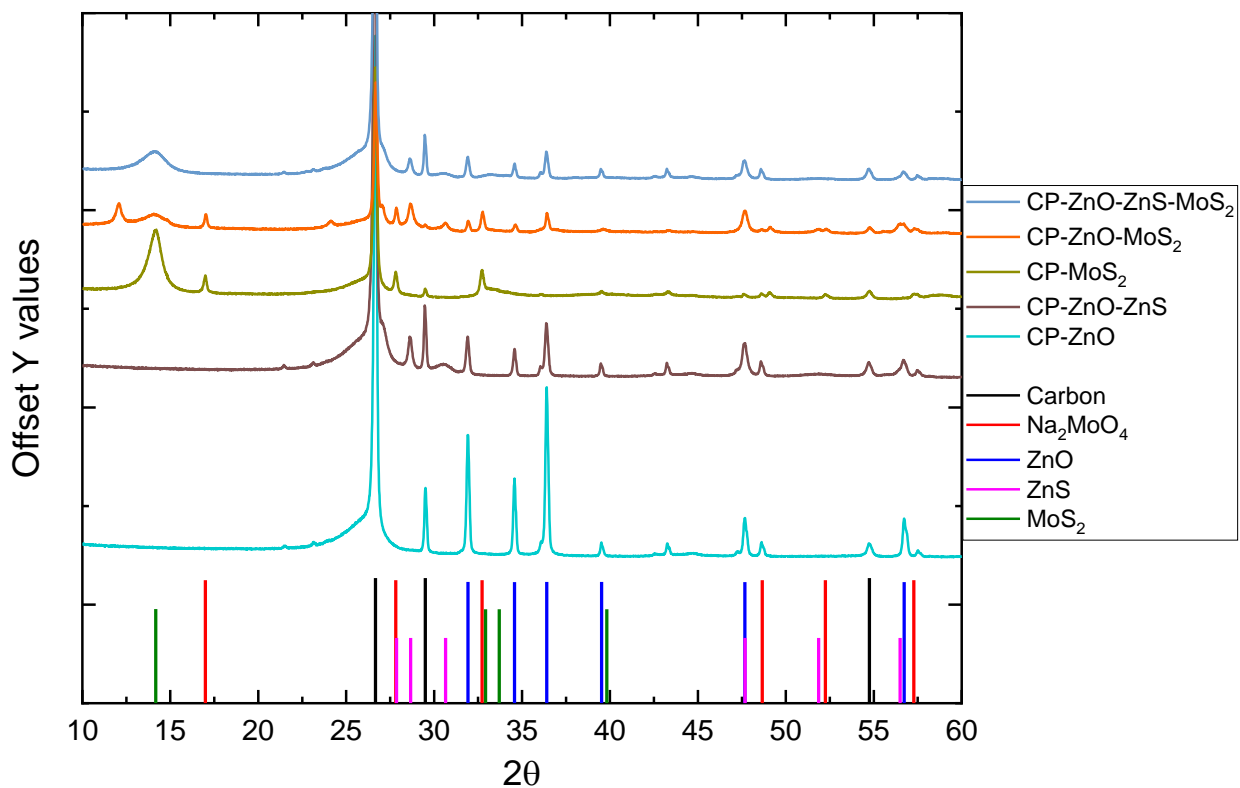

Figure 7: XRD pattern of the various nanowire electrocatalysts

The zoomed in region between 10° - 20° can be seen in Figure 8. As can be seen, there is an extra peak at 2θ = 12.08° and a small hump at 14.83° in the ZnO-MoS$_2$ sample. This has been duplicated and therefore is unlikely an outlier. Regarding the peak at 12.08°, we hypothesize that this is associated to an expanded c-axis MoS$_2$, a phenomenon that is reported in literature when MoS$_2$ is intercalated with various ions, or the formation of the 1T phase. However, applying Bragg's Law, the interlayer spacing is expanded by 104 pm from normal MoS$_2$, suggesting that oxygen is the intercalant given the atomic radius of oxygen (~50 pm). This has been reported in literature and is consistent with our findings.[26,27] This would also help explain why this peak is not present in the sulfurized sample, as the excess sulfur environment during the anneal would preferentially bond and prohibit the oxygen from diffusing between the MoS$_2$ sheets. It is interesting to note the sharpness of the 12.08° peak. The peak at 14.83° would suggest that the interlayer spacing has shrunk by ~26.2 pm. This has been attributed to the interface between ZnO and MoS$_2$ cause by lattice mismatch.[28,29] Both the ZnO-MoS$_2$ and ZnO-ZnS-MoS$_2$ samples have significant broadening and decreased intensity over the reference MoS$_2$, suggesting a decrease in crystallite size.

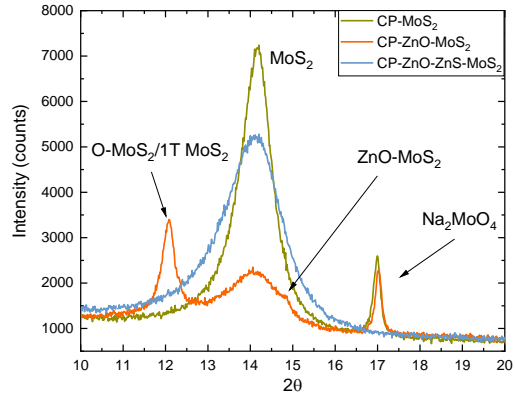

Figure 8: XRD patterns of the primary MoS$_2$ region of the MoS$_2$ containing nanowires.

A closer look at the Raman data from Figure 5 is plotted below in Figure 9. In the spectrum, the peaks at roughly 384 cm$^{-1}$ and 407 cm$^{-1}$ are attributed to the in-plane E$_{2g}$ mode and the out-of-plane A$_{1g}$ mode, respectively.[30] In addition, the forbidden E$_{1g}$ mode is present in all samples, indicating broken symmetry and is associated with edge species.[31] In addition to the Raman modes for 2H-MoS$_2$, distinct phonon modes at 154 (J$_1$), 219 (J$_2$), and 327 cm$^{-1}$ (J$_3$) are noted. These modes are characteristic of the metallic 1T-MoS$_2$, a reported side effect of the intercalation of ions.[32–34] This further supports the XRD results found in Figure 8. Additionally, the primary modes of A$_{1g}$ and E$^1_{2g}$ have shifted by ~2.2 cm$^{-1}$, indicating that the MoS$_2$ is under tensile stress.[35,36] Finally, the E$_2$ mode of ZnO and the E$_1$ mode can be seen at 439 and 349 cm$^{-1}$, respectively.

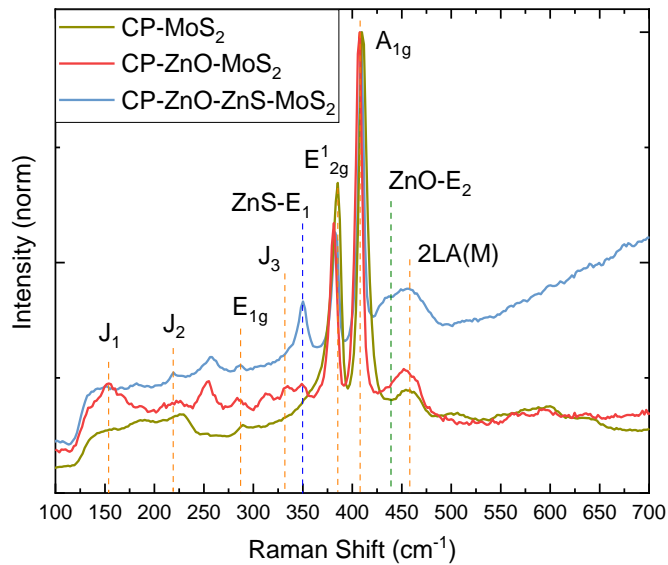

Figure 9: Raman of CP-MoS$_2$, CP-ZnO-MoS$_2$, and CP-ZnO-ZnS-MoS$_2$

### 3.1.3.3 Chemical Properties of ZnO/MoS$_2$ and ZnO/ZnS/MoS$_2$

To understand the chemical structure of the various nanowires, XPS was conducted to understand the chemical structure of the various nanowire heterostructures and can be seen in Figure 10. Within the Zn 2p$_{3/2}$ spectra, two species are present, with ZnO being the lower binding energy species at ~1021.9 eV and ZnS present at higher binding energies at 1022.9 eV. Within the S 2p spectra, there are three primary species, those corresponding to MoS$_2$ at ~161.9 eV, ZnS at ~162.8 eV, and sulfate species at ~169.7 eV. The sulfur spectrum was deconvoluted with two doublets, with the constraints of an energy difference of 1.18 eV between the 2p$_{3/2}$ and 2p$_{1/2}$ and with an intensity ratio of 2:1. The 2p$_{3/2}$ and 2p$_{1/2}$ were summed together for ease of viewing. The Mo 3d spectra can be deconvoluted into distinctive doublet (3d$_{5/2}$ and 3d$_{3/2}$) peaks. The primary constituents are 1T-MoS$_2$, 2H-MoS$_2$, MoO$_2$, and MoO$_3$, with the 3d$_{5/2}$ located at 229 eV, 229.3 eV, 230.8 eV, and 232.4 eV respectively. The Mo 3d peaks were deconvoluted with a binding energy difference of 3.13 eV between the 3d$_{5/2}$ and 3d$_{3/2}$ and with a fixed intensity ratio of 3:2. The sulfur 2s is also present within the Mo 3d spectra and is fit using parameters from the S 2p spectra.

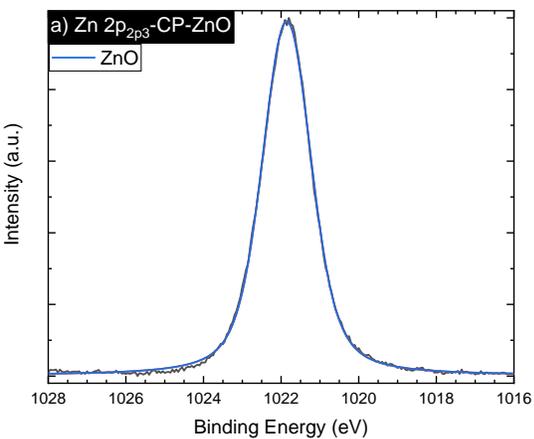
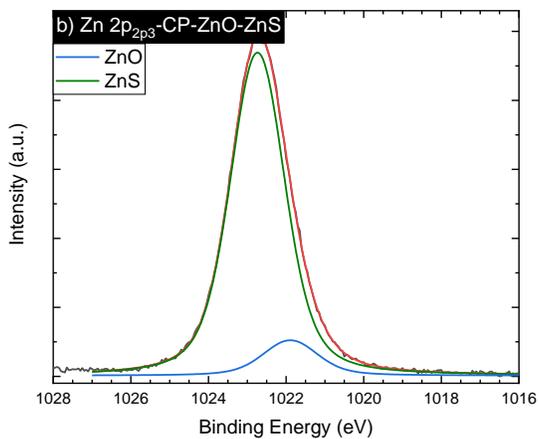
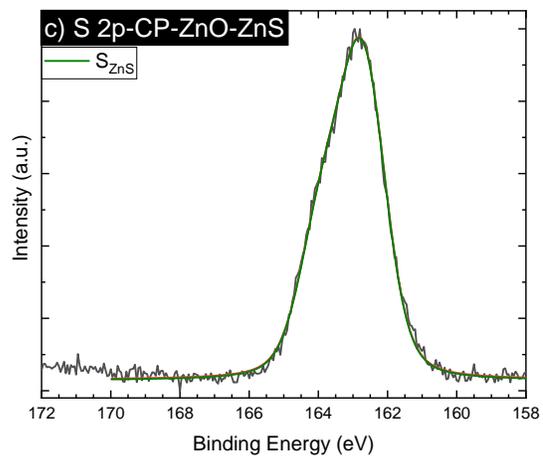
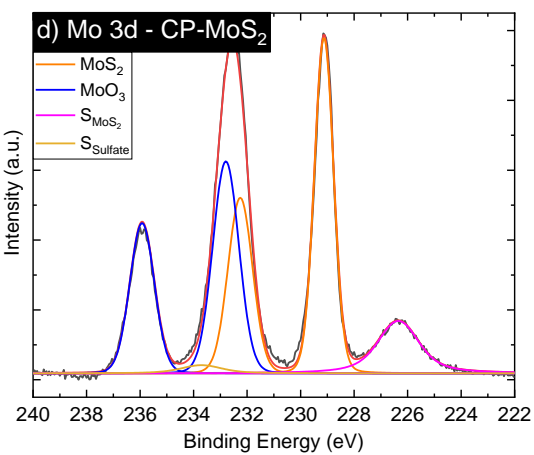
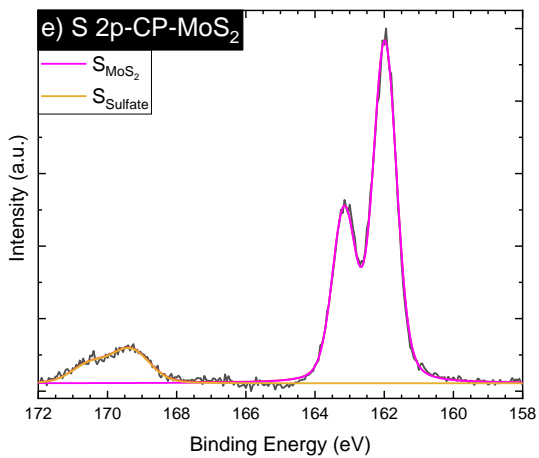
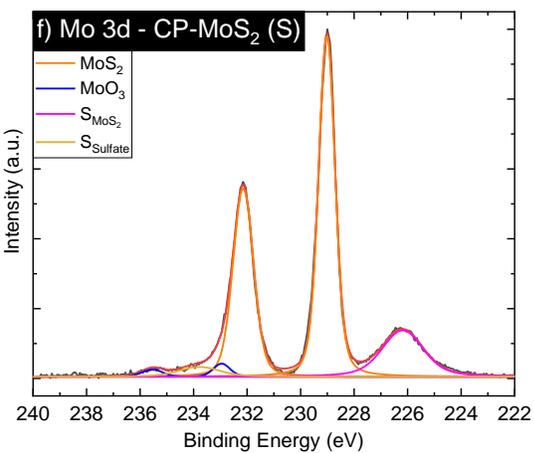
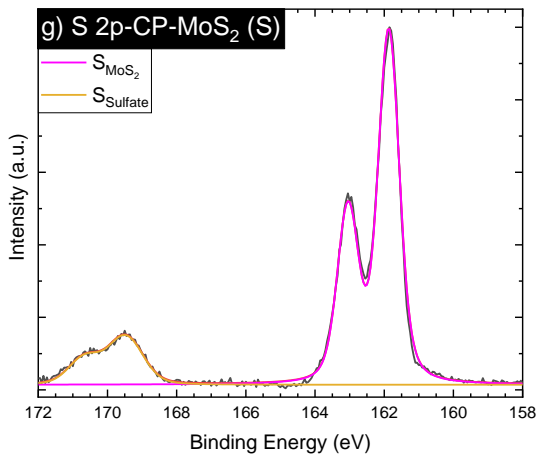

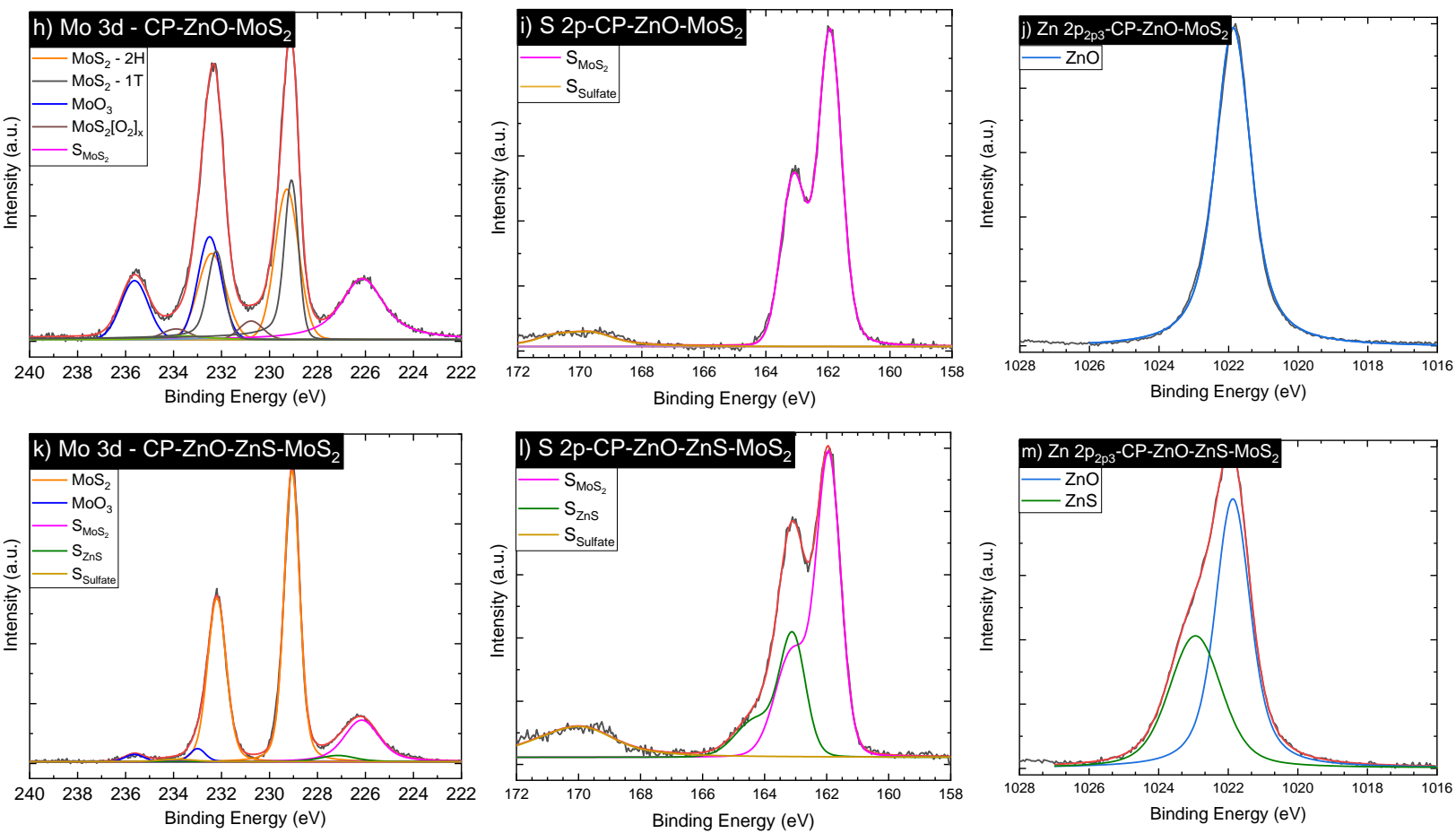

Figure 10: XPS of a) CP-ZnO nanowires, b,c) CP-ZnO-ZnS nanowires, d,e) CP-MoS$_2$ annealed with no sulfur, f,g) CP-MoS$_2$ annealed with sulfur, h,I,j) CP-ZnO-MoS$_2$ nanowires, and k,l,m) CP-ZnO-ZnS-MoS$_2$ nanowires

Starting from the top of Figure 10, the main differences between the CP-ZnO and CP-ZnO-ZnS is that only ZnO is present in the ZnO sample, whereas the spectra is dominated by ZnS in the CP-ZnO-ZnS sample. Moving down to Figure 10 d-g, which are the CP-MoS$_2$ samples without and with sulfur annealing, the primary difference is the large presence of the MoO$_3$ oxide phase in the MoS$_2$ sample that was annealed without any sulfur. This is expected as the unsaturated bonds would oxidize quickly to form MoO$_3$. Comparing the CP-ZnO-MoS$_2$ and CP-ZnO-ZnS-MoS$_2$ samples, again a major difference is the presence of the MoO$_3$ oxide phase in the sample that didn't see sulfur annealing, CP-ZnO-MoS$_2$. In addition, the CP-ZnO-MoS$_2$ also exhibits the asymmetric peak characteristic of 1T MoS$_2$. While the peak positions are close, the spectra could not be fit without the addition of the asymmetric peak, further supporting the XRD and Raman results. Finally, a feature that has been attributed to MoS$_2$[O$_2$]$_x$ is present in the CP-ZnO-MoS$_2$ and not present in any of the other samples. This has been shown to be present in oxygen intercalated MoS$_2$ by forming a superlattice, again supporting the idea that oxygen is acting as the intercalant.[37–41] Regarding Zn, only ZnO is present in the CP-ZnO-MoS$_2$ sample whereas there is a mixture of ZnO and ZnS in the CP-ZnO-ZnS-MoS$_2$ sample.

The method by Kraut et al. is used to determine the band offsets using the equation [42]

$$\Delta V = \Delta E_{CL} + \left(E_{Mo\ 3d}^{MoS_2} - E_{VBM}^{MoS_2}\right) - \left(E_{Zn\ 2p}^{ZnO} - E_{VBM}^{ZnO}\right)$$

where $\Delta E_{CL} = E_{Zn\ 2p}^{ZnO/MoS_2} - E_{Mo\ 3d}^{ZnO/MoS_2}$ is the energy difference between Zn 2p and Mo 3d core levels (CL) obtained by measuring the ZnO/MoS$_2$ heterojunction. $E_{Mo\ 3d}^{MoS_2} - E_{VBM}^{MoS_2}$ and $E_{Zn\ 2p}^{ZnO} - E_{VBM}^{ZnO}$ are the energy differences between the Mo 3d, ZnO 2p CLs and the valence band maximums for the reference MoS$_2$ and ZnO samples. A method of linearly extrapolating the valence band spectra leading edge to the baseline is used to determine the positions of the valence band maximums in the valence band spectra.[42,43] The valence band spectra are shown below in Figure 11 and the values are listed in Table 1.

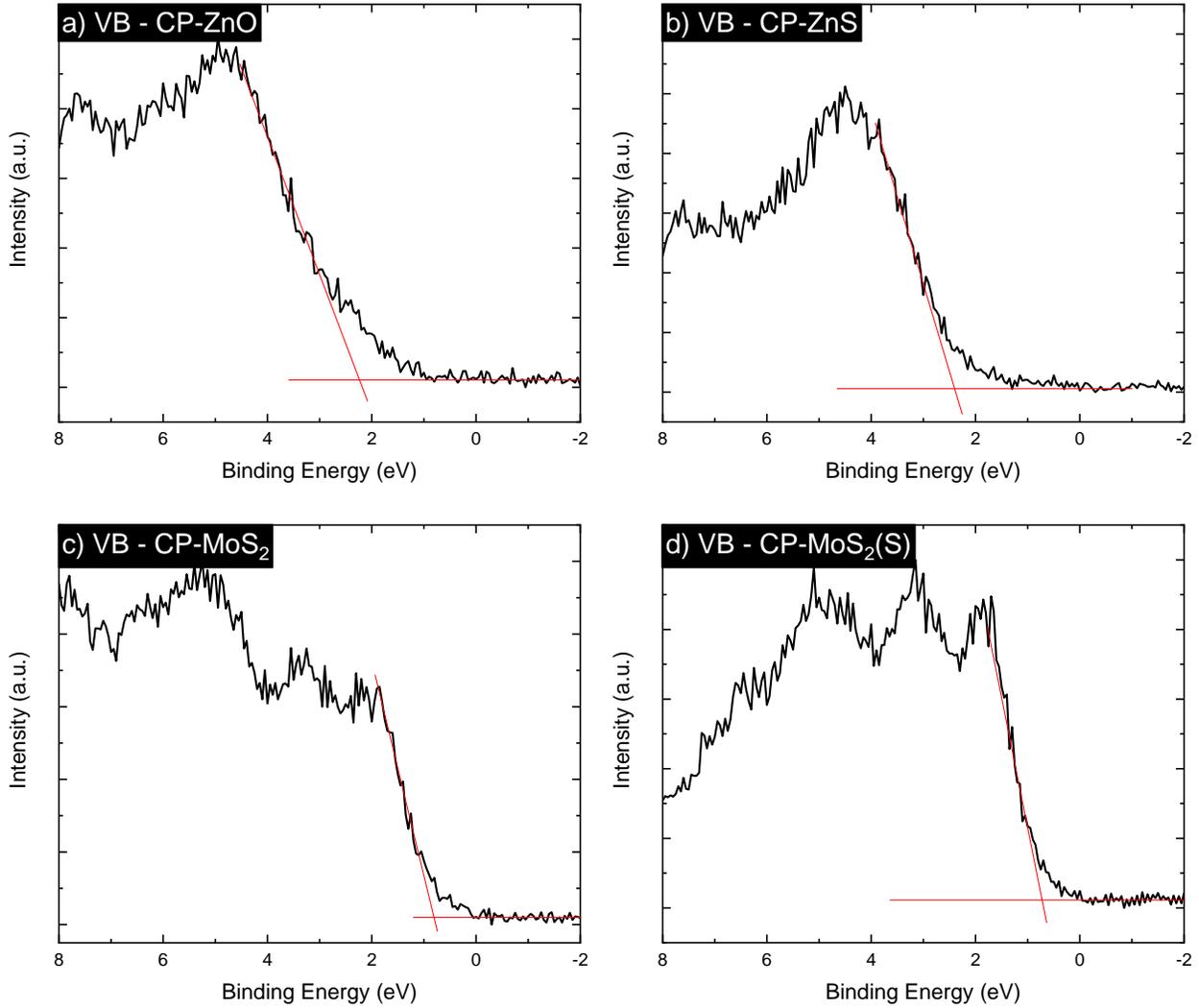

Figure 11: Valence band measurements of a) CP-ZnO nanowires b) completely sulfurized CP-ZnS nanowires c) CP-MoS$_2$ and d) CP-MoS$_2$ annealed in sulfur

Table 1: Binding Energies for the various chemical states

| Sample | State | Binding Energy (eV) |
| --- | --- | --- |
| CP-ZnO | Zn 2p | 1021.86 |
|  | VBM | 2.32 |
| CP-ZnS | Zn 2p | 1022.74 |
|  | VBM | 2.4 |
| CP-$MoS_2$ | Mo 3d | 229.126 |
|  | VBM | 0.8 |
| CP-$MoS_2$ (S) | Mo 3d | 229.017 |
|  | VBM | 0.72 |
| CP-ZnO-$MoS_2$ | Zn 2p | 1021.87 |
|  | Mo 3d | 229.276 |
| CP-ZnO-ZnS-$MoS_2$ | Zn 2p (ZnO) | 1021.83 |
|  | Zn 2p (ZnS) | 1022.99 |
|  | Mo 3d | 229.064 |

By following the above approach, a band diagram of the two studied nanowires can be constructed. However, we acknowledge that there is uncertainty for this approach. The method of linear extrapolation is commonly used in determining the VBM of semiconductors but is somewhat subjective as to where the linear line is drawn and its relationship with the Urbach tail. Another potential for systematic error is the spatially averaged nature of XPS. This leads to an uncertainty in exactly what is being analyzed, i.e. if there is any underlying, unbonded ZnO, ZnS, or $MoS_2$ that is not part of the nanowires, this would contribute to the signal. However, the complex geometry protruding from the surface, the surface sensitivity of the technique, and the minimal observed signal of carbon substrate, all suggest that any potential impact is likely minimal. The band diagrams can be seen in Figure 12.

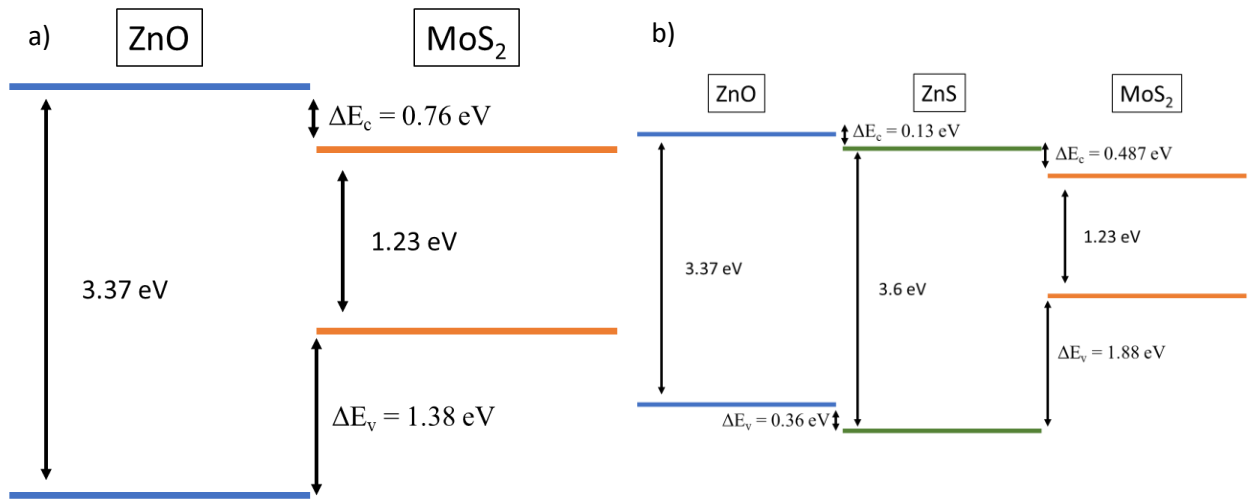

Figure 12: a) Band alignment diagram of CP-ZnO-MoS$_2$ nanowires and b) of the CP-ZnO-ZnS-MoS$_2$ nanowires

In the first of the examined structures, the CP-ZnO-MoS$_2$ nanowires exhibited a type-I band alignment where the MoS$_2$ is straddled by the ZnO. This is consistent with some of the findings mentioned earlier. [19,22] This would facilitate electron transfer from ZnO to MoS$_2$, but would not provide any benefit from charge separation as the holes would also flow from ZnO to MoS$_2$. In the CP-ZnO-ZnS-MoS$_2$ nanowire structure, a mixed type-II/type-I band alignment is noted. The interface between ZnO/ZnS exhibits a type-II band alignment where the electrons would flow from ZnO to ZnS but the holes would move from the ZnS to ZnO, providing some charge separation. The interface between ZnS and MoS$_2$ is a type-I band alignment where both the holes and electrons would accumulate in the MoS$_2$. This mixed band alignment is consistent with a previous report.[22] Overall, having the electrons accumulate in the MoS$_2$ where they can immediately perform water reduction at the surface would result in a net positive for catalysis.

**Electrochemical Properties of ZnO/MoS$_2$ and ZnO/ZnS/MoS$_2$**

To examine the impact the heterostructures have on the HER properties of the catalysts in acidic and alkaline conditions, polarization experiments were performed, the results of which are shown in Figure 13. A standard way of comparing catalysts is to determine the overpotential to reach 10 mA/cm$^2$. In both acidic and alkaline conditions, ZnO and ZnO-ZnS showed little activity, especially in alkaline conditions. The CP-MoS$_2$ sample annealed with the presence of sulfur performed notably better than the CP-MoS$_2$ without sulfur, likely due to the saturated bonds and the lack of the insulating MoO$_3$ phase. As can be seen, the CP-ZnO-ZnS-MoS$_2$ exhibited the lowest overpotential in both the acidic and alkaline conditions. CP-ZnO-MoS$_2$ demonstrated promising results in acidic conditions, performing greater than either ZnO or MoS$_2$ individually, but did not perform well in alkaline conditions. The performance and Tafel slopes are recorded in 2. The Tafel slope (b), i.e., the slope of the linear, low overpotential regime, is the overpotential required to increase the current density by one order of magnitude. As the Tafel slope is dictated by the reaction mechanism laid out in the experimental section, we can determine that the Volmer-Heyrovsky is the reaction mechanism for the ZnO-ZnS whereas the Volmer-Tafel mechanism is the reaction mechanism for all MoS$_2$ containing catalysts.[44]

Additionally, the slowest step for the MoS$_2$-containing catalysts is the Tafel reaction which involves the

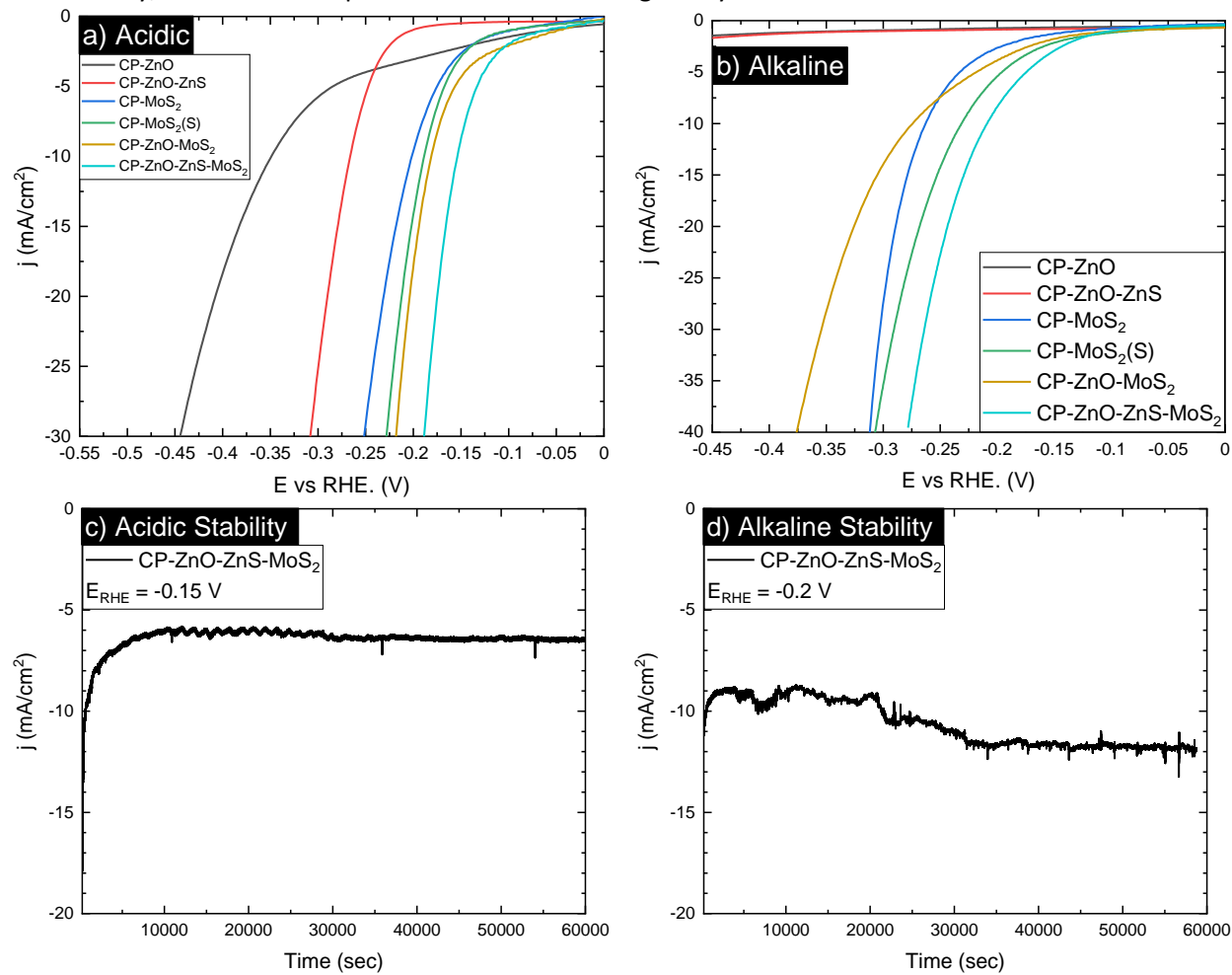

Figure 13 a) Polarization curves for samples in 0.5 M H$_2$SO$_4$ b) Polarization curves for samples in 1 M KOH. Stability measurements in c) acidic and d) alkaline conditions transfer of electrons along with the adsorption of the reacting species.

Table 2 Performance metrics for the various electrocatalysts

| Catalyst | Acidic | | Alkaline | | $C_{dl}$ (mF/cm$^2$) | Ref |
|---|---|---|---|---|---|---|
| | $\eta_{10}$ (mV) | Tafel Slope (mV dec$^{-1}$) | $\eta_{10}$ (mV) | Tafel Slope (mV dec$^{-1}$) | | |
| CP-ZnO | 349 | 127 | NA | NA | 8.45 | This work |
| CP-ZnO-ZnS | 266 | 60 | NA | NA | - | This work |
| CP-MoS$_2$ | 202 | 113 | 263 | 110 | 16.6 | This work |
| CP-MoS$_2$ (S) | 188 | 105 | 230 | 135 | - | This work |
| CP-ZnO-MoS$_2$ | 181 | 104 | 274 | 142 | 23.8 | This work |
| CP-ZnO-ZnS-MoS$_2$ | 154 | 103 | 209 | 128 | 25.0 | This work |
| | | | | | | |
| ZnO-MoS$_2$ nanocomposite | 255 | 56 | - | - | 3.56 | 45 |
| Nanosized ZnO coated MoS$_2$ | - | - | 550 | 192 | 1.78 | 46 |
| MoS$_2$/ZnS | 271 | 94.2 | - | - | - | 47 |
| ZnS-ZnO-MoS$_2$/Ti$_3$C$_2$T$_x$ | 327.6 | 79.5 | - | - | 7.3 | 48 |
| Nb$_2$/MoS$_2$-CNF | 180 | 29.5 | 125 | 33 | - | 49 |
| ZnS@MoS$_2$ | 194 | 73 | - | - | 2.07 | 50 |
| ZnS@MoS$_2$ Core/shell | 106 | 74 | - | - | 6.68 | 51 |

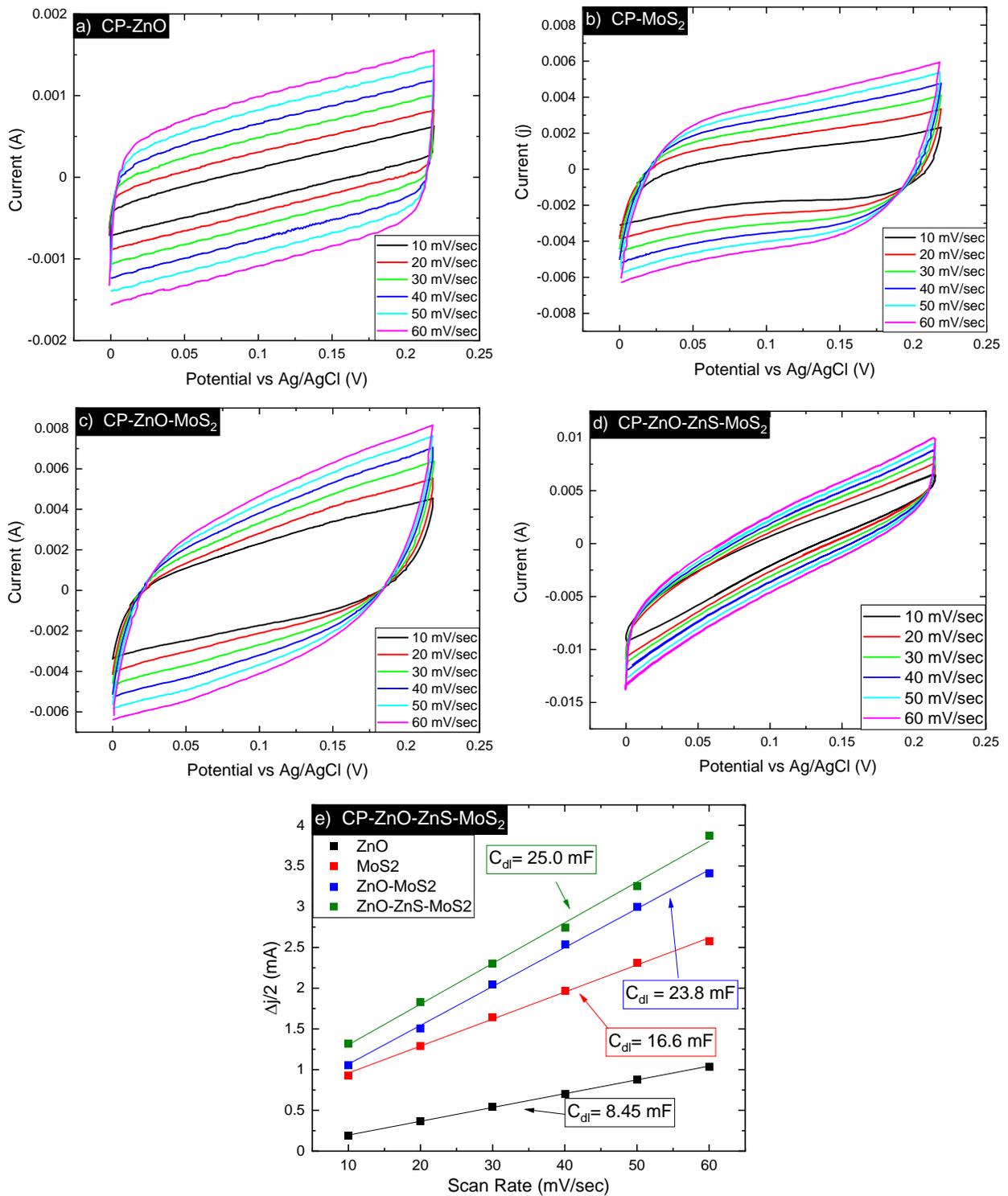

Figure 14: a) cyclic voltammograms of CP-ZnO b) cyclic voltammograms of CP-MoS$_2$ c) cyclic voltammograms of CP-ZnO-MoS$_2$ d) cyclic voltammograms of CP-ZnO-ZnS-MoS$_2$ and e) plot of Δj/2 as a function of scan rate

To complement the polarization curves, the double layer capacitance ($C_{dl}$) of select catalysts was conducted to give insight into the electrochemically active surface area (ECSA). This is useful when working with substrates that have changing complex geometries. ECSA can be determined from the $C_{dl}$ and the specific capacitance ($C_s$) of any investigated electrode material according to the equation[52]

$$ECSA = \frac{C_{dl}}{C_s}$$

Determination of $C_s$ has been identified as one of the main challenges with this technique and it has been common practice to use a single $C_s$ value to obtain the ECSA value, regardless of the material's nature. This has led to significant deviations in reported literature.[52,53] It is still an ongoing discussion within the community for how to accurately determine this value, however the comparison between $C_{dl}$ is still useful for comparisons between the same material and within the same study.[54]

A common method for obtaining the $C_{dl}$ consists of recording cyclic voltammograms at various scan rates within a potential region where no redox reactions take place as shown in Figure 14a-d. This works on the basis of when an electrode surface is subjected to a voltage ramp, a steady-state capacitive current is observed in a short time ($i_C$) if the only process is the charging of the double layer, i.e. the movement of ions on either side of the electrode/electrolyte interface. For ideal capacitors, $i_C$ is related to the capacitance (C) and to the scan rate (v) as shown below

$$i_C = v * C$$

The $i_C$ for the anodic and cathodic scans is the difference between them at the midpoint which is then plotted as a function of scan rates, as shown in Figure 14e. $C_{dl}$ is extracted by taking the slope of the resulting plot. $C_{dl}$ can also be obtained from EIS, however the difference between the techniques has been shown to be within 5%.[54–56] As can be seen in Figure 14e, all samples show significantly greater $C_{dl}$ over the bare ZnO-NWs even though the ZnO NWs have a significant surface area advantage over CP-$MoS_2$. This compliments our results earlier of CP-ZnO having poor catalytic performance. The $C_{dl}$ of CP-$MoS_2$, CP-ZnO-$MoS_2$, and CP-ZnO-ZnS-$MoS_2$, are somewhat similar however the nanowire structures exhibit a slightly higher $C_{dl}$, suggesting that there is an increase in ECSA attributable to the increased surface area that the ZnO nanowire structure provides. This increase in surface area compliments the formation of the heterojunctions as shown in Figure 12 where the lower CBM of the $MoS_2$ relative to ZnO and to ZnO-ZnS. This facilitates the more effective transfer of electrons to the increased surface area of the $MoS_2$ to enhance the active sites of $MoS_2$ to undergo the hydrogen evolution reaction. While it is not possible to completely separate the two effects here, given that the surface area between the ZnO-$MoS_2$ and ZnO-ZnS-$MoS_2$ structures are quite similar, and that even without the presence of the very active 1T phase, the ZnO-ZnS-$MoS_2$ is more active, the band alignment appears to be a large contributor to the increased performance.

**Conclusions**

This study developed a facile and scalable 3-step hydrothermal, electrodeposition, and annealing technique to synthesize a variety of nanowire heterostructures. The heterojunction catalysts of CP-ZnO-$MoS_2$ and CP-ZnO-ZnS-$MoS_2$ both saw an increase in catalytic activity in the acidic regime, with overpotentials to reach 10 mA/$cm^2$ of 181 mV and 154 mV respectively, over their constituent materials.

The CP-ZnO-ZnS-MoS$_2$ catalysts also saw an increase in catalytic activity in the alkaline conditions, requiring only 209 mV to reach 10 mA/cm$^2$. This has been attributed primarily to the synergistic properties of the band alignment structure and the increased surface area afforded by the nanowire structure. The 1T phase was also noted in the CP-ZnO-MoS$_2$ sample, suggesting that this may also play a role in increasing the catalytic activity of the sample. Overall, the synthesized nanowire heterostructures were found to be high-performing and gives insight into how the changing band alignment and morphology can play a significant role in increasing the catalytic performance towards HER.